\documentclass[12pt,preprint]{aastex}
\received{}
\begin{document}
\newcommand{\caii}{Ca{\,\sc ii}}
\newcommand{\msig}{$\mbh$--$\sigma$}
\newcommand{\kms}{{\rm km\;s^{-1}}}
\newcommand{\msol}{{\rm M_{\odot}}}
\newcommand{\mbh}{M_{BH}}
\newcommand{\ledd}{L_{Edd}}
\newcommand{\mdot}{\dot{m}}
\newcommand{\mdotedd}{\dot{m}_{Edd}}
\newcommand{\rg}{r_{g}}
\newcommand{\ang}{\rm \AA}

\title{Black Hole Masses of Active Galaxies with Double-Peaked
Balmer Emission Lines\altaffilmark{1}}

\shorttitle{Black Hole Masses of Double-Peaked Emitters}

\author{Karen T. Lewis\altaffilmark{2} and Michael Eracleous}

\affil{Department of Astronomy and Astrophysics, The
Pennsylvania State University, 525 Davey Laboratory, University Park,
PA 16802.}

\altaffiltext{1}{Based on observations carried out at Cerro Tololo
Interamerican Observatory, which is operated by AURA, Inc. under a
cooperative agreement with the National Science Foundation.}

\altaffiltext{2}{Present Address: NASA/Goddard Space Flight Center,
Code 662, Greenbelt, MD 20771, e-mail: {\tt
ktlewis@milkyway.gsfc.nasa.gov}}

\begin{abstract}
We have obtained near-IR spectra of five Active Galactic Nuclei that
exhibit double-peaked Balmer Emission Lines (NGC~1097, Pictor~A,
PKS~0921--213, 1E~0450.30--1817, and IRAS~0236.6--3101). The stellar
velocity dispersions of the host galaxies were measured from the
\caii~$\lambda\lambda 8494,8542,8662~\ang$ absorption lines and were
found to range from 140 to 200~$\kms$. Using the well-known
correlation between the black hole mass and the stellar velocity
dispersion, the black hole masses in these galaxies were estimated to
range from $4\times 10^{7}$ to $1.2\times 10^{8}~\msol$. We supplement
the observations presented here with estimates of the black holes
masses for five additional double-peaked emitters (Arp~102B, 3C~390.3,
NGC~4579, NGC~4203, and M81) obtained by other authors using similar
methods. Using these black hole masses, we infer the ratio of the
bolometric luminosity to the Eddington luminosity, ($L_{bol}/\ledd$).
We find that two objects (Pictor~A and PKS~0921--213) have
$L_{bol}/\ledd \sim\;$0.2, whereas the other objects have
$L_{bol}/\ledd \lesssim 10^{-2}$ (nearby, low-luminosity
double-peaked emitters are the most extreme, with $L_{bol}/\ledd
\lesssim 10^{-4}$).  The physical time scales in the outer regions of
the accretion disks (at $r\sim 10^3\; GM/c^2$) in these objects were
also estimated and range from a few months, for the dynamical
time scale, to several decades for the sound crossing time scale.  The
profile variability in these objects are typically an order of
magnitude longer than the dynamical time, but we note that variability
occurring on the dynamical time scale has not been ruled out by the
observations.

\end{abstract}


\section{INTRODUCTION}\label{mass_intro}

During the past two decades, the number of Active Galactic Nuclei
(AGN) known to exhibit broad, double-peaked Balmer emission lines has
been steadily increasing. These lines were first observed in the
broad-line radio galaxies (BLRGs) Arp~102B \citep*{ssk83,ch89,chf89},
3C~390.3 \citep{o87, p88}, and 3C~332 \citep{h90} and are reminiscent
of the double-peaked emission lines observed in Cataclysmic Variables,
where they are regarded as the kinematic signature of the accretion
disk \citep[e.g.,][]{hm86}. Thus it was suggested by the above authors
that the double-peaked emission lines originate in the accretion disk
that fuels the AGN.  A systematic survey of $z<0.4$ BLRGs and
radio-loud quasars was undertaken later on and it was found that
$\sim$~20\% of the observed objects exhibited double-peaked Balmer
lines \citep{eh94,eh03}.  More recently, \citet{s03} found that 3\% of
the $z<0.332$ AGNs in the Sloan Digital Sky Survey \citep[SDSS;
][]{y00} are double-peaked
emitters. \footnotemark\footnotetext{\citet{s03} finds that
double-peaked emitters are preferentially radio-loud, thus the larger
sample fraction among BLRGs is expected. See \citet{s03} for a
description of the differences between these two samples.}
Double-peaked emission lines have also been observed in Low Ionization
Nuclear Emission Line Regions \citep[LINERs; ][]{h80} including
NGC~1097 \citep{sb93}, M81 \citep{b96}, NGC~4203 \citep{s00}, NGC~4450
\citep{h00}, and NGC~4579 \citep{b01}. Given the difficulty of
detecting broad emission lines in LINERs \citep{ho97} it is likely
that a larger number of LINERs exhibit double-peaked emission
lines. Although the known double-peaked emitters represent a small
fraction of the total AGN population, they are an important and
intriguing class of objects that deserve further study.

As described in detail by \citet{eh03}, observational tests and basic
physical considerations strongly suggest that the double-peaked Balmer
lines originate in the outer regions of the accretion disk at
distances from the black hole of hundreds to thousands of
gravitational radii ($\rg$~=~$G\mbh/c^{2}$, where $\mbh$ is the mass
of the black hole). The {\it profiles} of the double-peaked emission
lines are observed to vary on time scales of years to decades and show
occasional reversals in the relative strength of the red and blue
peaks \citep*[for examples,
see][]{vz91,zvg91,g99,shap01,sps00,serg02,sb03,suvi04,l04a,l04b,lewis05,gezari05}.
Intensive monitoring of 3C~390.3 shows that the time lag between
variations in the continuum and the optical broad line is $\sim$ 20
days \citep{d98,serg02} thus the profile variations occur on much
longer time scales than the light-crossing time and seem to be
unrelated to fluctuations in the ionizing continuum flux that take
place on time scales of a few days to a few weeks.  A similar
effect is also observed in systematically monitored Seyfert galaxies
\citep[e.g.,][]{wp96,k97}. The profile variability is likely to be a
manifestation of {\it physical} changes in the outer disk and the
profile variability is thus a powerful tool that can be used to test
models for dynamical behavior in accretion disks. Therefore, in the
early 1990s a campaign was undertaken to systematically monitor a set
of $\sim 20$ double-peaked emitters, including BLRGs from the
\citet{eh94} sample and several LINERs.

\citet{eh03} and \citet{s03} found that the profiles of approximately
60\% and 40\%, respectively, of the double-peaked emitters from their
samples can be fitted by a simple, axisymmetric accretion disk model.
However the long-term profile variability and the fact that in some
objects the red peak is stronger than the blue peak (in contrast to
the expectations of relativistic Doppler boosting) require the use of
more general, non-axisymmetric models.  Some of the models that have
been successfully employed to fit observed profiles include circular
disks with orbiting bright spots \citep[Arp~102B and 3C~390.3;
][]{ne97,sps00,zvg91} or spiral emissivity perturbations
\citep[3C~390.3, 3C~332, and NGC~1097; ][]{g99,sb03} and precessing
elliptical disks \citep[NGC~1097, ][]{e95,sb95,sb97,sb03}.

To successfully interpret and model the long-term profile variability,
it is not sufficient to reproduce the observed sequence of profiles;
the variability must occur on a physical time scale that is consistent
with the chosen model. The time scales of interest are the light
crossing, dynamical, thermal, and sound crossing times, which are set
by the black hole mass ($\mbh$) and are given by \citet*{fkrbook}:
\begin{eqnarray}
 \tau_{\ell} & = & 6\; M_8\;\xi_3 \, {\rm days} \label{tl}\\
 \tau_{dyn} & = &  30\; \tau_{\ell}\;\xi_3^{1/2}\; = \;6\; M_8\;\xi_3^{3/2} \, {\rm months} \label{tdyn}\\
\tau_{th}  & = & \tau_{dyn}/\alpha\; = \;5\; M_8\;\xi_3^{3/2} \, {\rm years}\\
\tau_{s}   & = & 70\; M_8 \;\xi_3\; T_5^{-1/2} \, {\rm years} \label{ts}
\end{eqnarray}
\noindent where $M_{8}=\mbh/10^8\;\msol$, $\xi_{3} = 
r/10^{3}\;\rg$, $ T_{5}= T/10^{5}\;$K, and $\alpha~(\sim 0.1)$
is the Shakura-Sunyaev viscosity parameter \citep{ss73}. The above
models all predict variability on different time scales. For example
matter embedded in the disk orbits over $\tau_{dyn}$, thermal
instabilities will dissipate over $\tau_{th}$, density perturbations,
such as a spiral wave, precess on time scales that are an order of
magnitude longer than $\tau_{dyn}$ up to $\tau_{s}$, and an elliptical
disk will precess over even longer time scales.

Many of the models described above yield strikingly similar sequences
of profiles. In order to determine which physical mechanism is
responsible for the profile variability, it is necessary to connect
the observed variability time scale (in years) with one of the above
{\it physical} time scales. Using an estimate of the black hole mass in
NGC~1097, \citet{sb03} estimated the physical time scales in the outer
accretion disk which favored a spiral arm model over an elliptical
disk model for this object. However, in order to discriminate between
the above time scales, the black hole mass must be known to better than
an order of magnitude and preferably to better than a factor of
three.

Another motivation for accurately estimating the black hole masses in
the double-peaked emitters is to determine the accretion rate,
relative to the Eddington accretion rate ($\mdot/\mdotedd$), which is
a diagnostic of the structure of the inner accretion flow. Because the
local energy dissipation in the accretion disk is not always
sufficient to power the observed H$\alpha$ luminosity, \citet{ch89}
proposed that the inner accretion flow in double-peaked emitters has
the form of an ion torus \citep{r82}, or similar vertically extended,
radiatively inefficient flow. This vertically extended structure would
be able to illuminate the outer accretion disk and power the observed
emission lines.  For the well-studied \citet{eh94} sample there is
considerable indirect observational support of this hypothesis,
discussed fully in \citet{eh03}, which is underscored by the growing
number of LINERs known to sport double-peaked emission lines. The
assumption that the inner accretion flow is radiatively inefficient is
fundamental to the current ideas for the formation of the
double-peaked emission lines and also the relationship between
double-peaked emitters and the AGN population in general.  Thus it is
extremely important to test this hypothesis more directly by obtaining
estimates of the black hole masses, and thus $\mdot/\mdotedd$, for
double-peaked emitters.

Most of the double-peaked emitters are too distant to obtain black
hole masses directly via spatially resolved stellar and gas
kinematics. However, it is possible to determine the black hole masses
indirectly through the well-known correlation between $\mbh$ and the
{\it stellar} velocity dispersion ($\sigma$) measured on the scale of
the effective radius of the bulge \citep{fm00,g00,trem02}. Therefore
we have begun a program to measure the stellar velocity dispersions in
double-peaked emitters using the \caii~$\lambda 8594,8542,8662$
triplet and present here the results for five objects. After obtaining
estimates of the black hole masses via the \msig~relationship, the
physical time scales in the outer accretion disk and $\mdot/\mdotedd$
can be inferred for each object.

Here, we focus specifically on measurements of the velocity dispersion
in these double-peaked emitters to enable determinations of black
hole masses. Although the time scales derived here are extremely
valuable in modeling the long-term profile variability in these
objects, a detailed discussion of the profile modeling is deferred to
forthcoming papers. Preliminary results on modeling the profile
variability (with the help of the black hole masses obtained here) can
be found in \citet{lewis05} and \citet{gezari05}.

This paper is organized as follows. In \S\ref{mass_data}, we describe
the target selection, observations and data reductions. The analysis
of the data, including an examination of the various sources of error,
are described in \S\ref{mass_analysis} and the results and their
implications are presented and discussed in
\S\ref{mass_discussion}. In Appendix \ref{mass_ap1}, we present a
detailed description of the procedure used to correct the telluric
water vapor absorptions lines in these spectra. Throughout the paper,
we assume a WMAP cosmology \citep[$H_{\rm 0}=70~{\rm
km~s^{-1}~Mpc^{-1}}$, $\Omega_{\rm M}=0.27$,
$\Omega_{\Lambda}=0.73$;][]{wmap}.

\section{SAMPLE SELECTION, OBSERVATIONS, AND DATA REDUCTION}\label{mass_data}

Our primary motivation for obtaining robust black hole masses is to
assist with modeling and interpreting the long-term profile
variability of the double-peaked emitters. Therefore, we selected the
targets that were part of the long-term monitoring campaign and have
shown interesting variability.  Absorption due to telluric water vapor
becomes severe at wavelengths longer than 9200~$\ang$, so only objects
with $z<0.062$ were selected. Finally, we only selected objects with
declinations less than $-5^{\rm \circ}$. Five objects, NGC~1097,
Pictor~A, PKS~0921--213, 1E~0450.3--1817, and IRAS~0236.6--3101, met
these criteria, and the properties of these objects are given in Table
\ref{gal_obs}.  All of our targets are hosted by elliptical galaxies
or early-type spirals, whose stellar spectra in the \caii\ region are
dominated by G and K giants \citep{w94}. Therefore we observed
numerous G and K giant stars, with spectral types ranging from K5 to
G6, to serve as stellar templates.  The stars used in the final fits
are listed in Table \ref{star_obs}.

The spectra were obtained on 2003 December 4--8 using the RC
spectrograph on the 4m Blanco telescope at the Cerro-Tololo
Ineteramerican Observatory.  The G~380 grating was used in conjunction
with the RG 610 order-separating filter and the spectra covered the
range from 7690 to 9350~$\ang$. The slit had a width of
$1.\!\!^{\prime\prime}$33 and was oriented east to west.  The
resulting spectral resolution was 1.35~$\ang$ FWHM, as measured from
the arc lamp spectra, corresponding to a velocity resolution of $\sim
50\;\kms$.  The galaxies and the template stars were observed at an
airmass of less than 1.13 and differential atmospheric refraction was
not significant over the small wavelength interval of interest.The
atmospheric seeing during the observations of the galaxies was less
than $1.\!\!^{\prime\prime}5$ and frequently less than
$1.\!\!^{\prime\prime}0$, although some of the template stars were
observed with seeing as large as $2.\!\!^{\prime\prime}0$.

A set of bias frames and HeNeAr comparison lamp spectra were taken at
the beginning and end of each night and quartz flats were taken
immediately preceding or following each galaxy exposure. Rapidly rotating B
stars (with rotational velocities in excess of 200~$\kms$) were
observed periodically throughout the night at similar airmass as the
galaxies and template stars and were used to correct for the deep
telluric water vapor lines at wavelengths longer than 8900~$\ang$ (in
the galaxies) and shorter than 8400~$\ang$ (in the template
stars). The exposure times for the galaxies are listed in Table
\ref{gal_obs}.

The primary data reductions --- the bias level correction, flat
fielding, sky subtraction, removal of bad columns and cosmic rays,
extraction of the spectra, and wavelength calibration --- were
performed using the Image Reduction and Analysis Facility (IRAF)
\footnotemark\footnotetext{IRAF is distributed by the National Optical
Astronomy Observatories, which are operated by the Association of
Universities for Research in Astronomy, Inc., under cooperative
agreement with the National Science Foundation.}.  The night sky
emission was estimated by fitting a 3rd order spline along the spatial
direction to two intervals (with widths of 10--20 pixels) located on
either side of the spectral extraction region. In most objects, with
the exception of Pictor~A, the Ca II region was not heavily
contaminated by night sky emission lines. The issue of sky subtraction
will be discussed further in \S\ref{fit_method}.  The difference in
dispersion solutions from one night to the next was less than 0.1\%,
so the same dispersion solution was applied to the spectra from all
five nights. It was not possible to obtain comparison lamp spectra
before or after each galaxy spectrum was taken. However, measurements
of the night sky lines, which are largely unblended at this
resolution, were used to fine-tune the wavelength solution for the
galaxy spectra before they were combined; this ensured that the
combined spectra were not artificially broadened. The applied shifts
were generally less than 1$\;\ang$. The error spectrum was calculated
by adding in quadrature the Poisson noise in the spectra of the night
sky and the object spectrum (prior to the removal of cosmic rays and
bad columns).

None of the host galaxies in our sample have a reported value of the
effective radius, and in all cases except NGC~1097 and Pictor~A,
suitable images were not available from which to determine the
effective radius.  To obtain a rough estimate of the effective radius,
the spectra were collapsed along the spectral direction to obtain a
spatial profile, which was then fitted with the sum of a de
Vaucouleurs (R$^{1/4}$) profile and an AGN point source convolved
with a Gaussian, and background. The effective radius determined for
Pictor~A from a ground-based image was in agreement with that obtained
from the spatial profile. There was considerable uncertainty in the
effective radii, however, and we chose to extract the spectra using
the smallest effective radius allowed by the data (see also the
discussion in \S\ref{mass_error_source}). In the case of NGC~1097,
which is a spiral galaxy with a nuclear star forming ring, the
extraction radius was chosen to lie just within the star forming ring
(see \S\ref{mass_notes} for more details). The extraction radii used
for the galaxies are listed in Table \ref{gal_obs}.

The removal of the telluric water vapor lines at wavelengths longer
than 8900~$\ang$ was an essential step in the data reductions, because
many of the \caii\ lines in the observed galaxies were redshifted into
this region of the spectrum. As described in detail in Appendix~A, a
telluric template of the atmospheric transmission was derived from the
spectrum of a rapidly rotating B-star. The spectra obtained from
individual exposures of the galaxies were divided by this telluric
template allowing for the possibility of a slight wavelength shift
between the template and galaxy spectra. Because the humidity was
quite low and stable throughout the run (25--35\%) and all of our
objects were observed at low airmass, we found that it was possible to
correct all of the galaxies with the same telluric template.

The spectra of the G and K giant stars as well as those from
individual exposures of the galaxies were normalized with a low order
polynomial. In most instances, the averaged galaxy spectra could be
successfully fitted without any further normalization. However, the
spectrum of NGC~1097 was re-normalized with a low-order polynomial in
the region of the \caii\ triplet.  The resulting galaxy spectra and a
representative template star are show in Figure~\ref{spectra}. The
average signal-to-noise ratios (S/N) of the galaxy spectra are listed
in Table~\ref{gal_obs}; the S/N of the stellar templates ranged
from 150 to 300.

\section{ANALYSIS AND RESULTS}\label{mass_analysis}

\subsection{Fitting Method}\label{fit_method}
The velocity dispersion in the host galaxies of these five AGNs were
determined by directly fitting the galaxy spectra with a model given by:
\begin{equation}
M(\lambda) = f\cdot T(\lambda) \otimes G(\lambda) + P(\lambda)
\end{equation}
where $T(\lambda)$ is a stellar template, $f$ ($<$ 1) is a dilution
factor, $G(\lambda)$ is a Gaussian with a dispersion $\sigma$,
$P(\lambda)$ is a low-order ($1^{\rm st}$ or $2^{\rm nd}$ order)
polynomial, and $\otimes$ denotes a convolution. The stellar template
was either an individual G or K giant star or a linear combination
of G and K giants \citep[with weights of 25\% and 75\% respectively;
following,][]{w94}. The fitting intervals were selected to exclude
emission lines (\ion{O}{1}~$\lambda$8446 and
[\ion{Fe}{2}]~$\lambda$8618) however it was not necessary to
explicitly exclude pixels contaminated by strong night sky emission
lines, bad columns, or cosmic rays; the large error bars on such data
points resulted in these data being effectively ignored in determining
the best fitting model.

To find the best fitting velocity dispersion and its uncertainty we
scanned the 2-dimensional parameter space defined by $\sigma$ and $f$
in small steps.  The coefficients of the low-order polynomial
describing the non-stellar continuum were evaluated at each grid point
by minimizing the $\chi^2$ statistic.  For all of the galaxies,
including NGC~1097, the value of $\chi^{2}$ per degree of freedom
($\chi^{2}_{\nu}$) for the best fit was typically $\chi^{2}_{\nu} \sim
3$. This large value of $\chi^{2}_{\nu}$ suggests that the error bars
on the flux density of the pixels in the spectra were underestimated
and in fact the root mean square (RMS) dispersion of the data in
featureless regions of the spectra was typically 1.5--1.9 times larger
than the formal error bars assigned to the individual pixels. We
attribute this increased scatter to imperfect subtraction of the
strong night-sky emission lines appearing throughout our spectral
range (this was caused by curvature of the sky lines along the
direction of the slit). We note that the discrepancy between the RMS
deviation and the formal error bars is only slightly greater at
wavelengths larger than 8900~$\ang$, where the telluric absorption
correction was performed. If we increased the error bars on the
spectral pixels by the above factor, the best fit models would have
had $\chi^{2}_{\nu} \sim 1$ in most cases.  Thus the values of
$\chi^{2}$ for each fit were rescaled such that $\chi^{2}_{\nu} \equiv
1$ and the 68\%-confidence error contour in $\sigma$ and $f$ was
defined by $\Delta\chi^{2} = 2.3$. This is equivalent to the practice
of rescaling the error bars on each pixel, adopted by \citet*{b02}.
The galaxy spectra and their best fitting models are shown in
Figure~\ref{spectra}.

The use of a direct fitting method for determining velocity
dispersions has been well tested by \citet{b02} and \citet{garcia05}
demonstrate that the results obtained with direct fitting are
consistent with those obtained through the Fourier Correlation
Quotient method \citep[see, e.g.][]{bender90}. Nevertheless, we also
conducted our own tests using simulated spectra that were generated
from a broadened K2 III stellar template (plus a featureless
continuum) to which random noise was added. In our simulated spectra,
the velocity dispersion ranged from 100 to 200$\;\kms$, the dilution
factor ($f$) varied from 0.6 to 1.0, the order of the featureless
continuum ranged from 0 to 2, and the S/N of the spectra ranged from
35 to 100. After generating ten realizations of each simulated
spectrum, we found that our direct fitting method successfully
recovered the input velocity dispersion within the 68\% error interval
determined from the fit. This was the case even when the order of the
polynomial representing the featureless continuum was forced to be
different from that used to generate the spectra. Of course, the error
bar on the best-fit stellar velocity dispersion was larger for spectra
with lower S/N and/or larger velocity dispersion and dilution
factor. We also tested our implementation of the direct fitting
method, which differs slightly from that used by \citet{b02}, by using
our code to fit the data for Arp~102B obtained by these
authors. Within the error bars, we obtain the same velocity dispersion
as these authors.

\subsection{Sources of Systematic Error}\label{mass_error_source}

There are some potential sources of systematic error which were not
accounted for by the error analysis described above.

\begin{description}

\item
{\it Extraction Size. --} The effective radii of the galaxies
determined here are quite uncertain, however we found that best-fit
velocity dispersion was not sensitive to the extraction radius. We
also note that the S/N of the spectra was not significantly affected
by the extraction radius, with the exception of Pictor~A which is
discussed further in \S\ref{mass_notes}.

\item
{\it Telluric absorption correction. --} Although considerable care
was taken to perform the telluric absorption correction, the telluric
template is most likely overestimated in the interval from 9195 to
9215$\ang$, as described in detail in Appendix
\ref{mass_ap1}. Consequently, the telluric correction in this interval
is not complete and the flux in the corrected galaxy spectra could be
underestimated by as much as 4\%. This artifact cannot be accounted
for in any statistical way, however when it is obvious in the spectra,
this interval is not included in the fit.

\item
{\it Template Mismatch. --} An important advantage of using the
\caii\ triplet to measure the stellar velocity dispersion is the fact
that the strength of the absorption lines is relatively insensitive to
the stellar population \citep{p78,d84}. We found that the RMS scatter
in the values of $\sigma$ obtained using the different templates (both
individual stars and linear combinations of stars) was only a
few~$\kms$. This is consistent with the RMS spread we obtain when
fitting high S/N simulated spectra with stellar templates that were
different than that used to generate the spectrum. This template
mismatch represents an additional error which is not accounted for by
our fitting procedure, which only explores a range in $\sigma$ and
$f$.  We treat this RMS spread as an additional error which we add in
quadrature to the 68\% error bar for NGC~1097, similarly to
\citet{b02}.  For the other galaxies the uncertainty due to template
mismatch was negligible compared to the uncertainty on the best-fit
and was not included.

\end{description}

\subsection{Notes on Individual Objects}\label{mass_notes}

\begin{description}

\item
{\it NGC 1097. --} The nuclear structure of the barred spiral galaxy
NGC~1097 is extremely complicated, as evidenced by $^{12}$CO
observations which suggest the presence of a cold nuclear disk being
fed by matter streaming along the bar \citep{emsellem01}. These
authors find that within the inner 5$\arcsec$, the velocity dispersion
ranges from 145~$\kms$ in the center to 220~$\kms$ at the inner edge
of the bar, as measured from the broadening of the CO band head. There
is a star forming ring extending from 5 to 10$\arcsec$, as mapped by
H$\alpha$ and radio emission \citep{hum87}, which is mostly excluded
in our data. However \citet{sb05} recently demonstrated that there is
a starburst within $0.\!\!^{\prime\prime}1$ (9~pc) of the central
black hole. Despite these difficulties, we obtained an excellent fit
to the data using a single K giant star, although the featureless
continuum is more complex and must be described with a 3$^{\rm rd}$
order polynomial. The interval from 8600-8640~$\ang$ (rest wavelength)
was excluded from the fit because contamination from [Fe II] emission
might be responsible for the poor fit to the data in this interval. If
this interval is included in the fit, the best-fit velocity dispersion
increases to 208$\pm5~\kms$.

\item
{\it Pictor A. --} The \caii\ triplet is located in the region of the
spectrum with the strongest night sky emission lines. Although the effective
radius was estimated to be 9$\arcsec$ (6.5~kpc), the use of a smaller
extraction radius ($4.\!\!^{\prime\prime}5$) allowed for a much improved
subtraction of the night sky emission, and consequently an increased
S/N and a decrease in the error bars. When a 9$\arcsec$ extraction
radius was used, the best-fit velocity dispersion was the same,
although the error bars were larger. Both \ion{O}{1} and [\ion{Fe}{2}]
emission lines are present in the spectrum, and the regions around
these lines were ignored in the fit.

\item
{\it PKS 0921--213. --} The \caii~$\lambda8662$ line is strongly
contaminated by several bad columns and as a result the blue side of
this line does not contribute to the determination of the velocity
dispersion.

 \item
{\it 1E~0450.3--1817. --} This object is extremely faint and the S/N
is much lower than for the other objects (S/N $\sim$ 15 per
pixel). Therefore the spectrum was smoothed by a 3-pixel wide boxcar
function in order to increase the S/N ($\sim$ 25 per pixel). The
best-fit velocity dispersion is the same as when the un-smoothed
spectrum is fitted, but the error bars are decreased by $\sim$30\%. We
have verified that for NGC~1097 and IRAS~0236.6--3101, neither the
best-fit velocity dispersion nor the uncertainty are changed when the
spectra are similarly smoothed.

\item
{\it IRAS~0236.6--3101. --} The \caii~$\lambda8662$ line has an
observed wavelength of 9202~$\ang$, which places this line in the
region of the telluric template that was underestimated, as described
in \S\ref{mass_error_source} and Appendix \ref{mass_ap1}. The
$\lambda8662$ line is deeper than would be expected based on the
strength of the two other \caii\ lines, and we chose to ignore this
line when performing the fit. If this line were included, the best-fit
value of $\sigma$ would increase to $\sim 200\;\kms$, however a model
with this large velocity dispersion is an extremely poor fit to the
$\lambda8542$ line. An \ion{O}{1} emission line is present, and
the interval around this line was excluded from the fit.

\end{description}

\section{DISCUSSION AND CONCLUSIONS}\label{mass_discussion}

Using the measurements of the velocity dispersion found above, we
determine the black hole masses, the Eddington ratios, and the
physical timescales in the accretion disk. All of these quantities are
given in Table \ref{results_tab}. The black hole masses were estimated
via the \msig~relationship \citep{trem02}, namely,
\begin{equation}
\log\left({M\over\msol}\right) = \alpha + \beta\;
\log\left({\sigma\over\sigma_0}\right)\; , 
\end{equation}
where $\alpha = 8.13 \pm 0.06$, $\beta = 4.02 \pm 0.32$, and $\sigma_0
= 200~\kms$.  The scatter in the \msig~relationship is not included in
the error, but the error bars on the coefficients are. For
completeness, we include in this table the double-peaked emitters with
stellar velocity dispersions reported in the literature, namely
Arp~102B, NGC~4203, and NGC~4579 \citep{b02} and 3C~390.3
\citep{nel04}. We also include M81, for which a mass estimate of
$\sim~6\times10^{7}~\msol$ is based on resolved stellar and kinematics
\citep[][respectively]{b00,dev03}. The inferred black hole masses for
these ten objects range from $4\times10^{7}~\msol$ to $5\times
10^{8}~\msol$.

We note the estimate of the black hole mass for NGC~4203 used here
($\mbh~\sim~6~\times10^{7}~\msol$) is in disagreement with the results
of \citet{s00} and \citet{sarzi01}, who placed an upper limit of
$6\,\times\,10^{6}~\msol$ on the black hole mass from spatially
resolved gas kinematics. As noted by \citet{s00}, the gas need not lie
in a flat disk and it could be subject to non-gravitational forces; in
the particular case of NGC~4203, the disk is is suspected to be warped
\citep{sarzi01}. Thus, we have chosen to use the mass obtained from
the stellar velocity dispersion.

The physical timescales in the outer accretion disk were computed
using equations~\ref{tdyn}--\ref{ts}, assuming that $\xi_{3} = 1$,
$\alpha = 0.1$ and $T_{5} = 1$. For these black hole masses, the
dynamical timescales ranges from a few months to one year, the thermal
timescales ranges from one year to a decade, and the sound crossing
timescales ranges from a decade to 100 years. The observed profile
variations occur on timescales of several years \citep[see, e.g.,
][]{lewis05,gezari05} for most objects, which is an order of magnitude
longer than the dynamical timescales found here.  This strongly
suggests that in general the profile variability might be due
predominantly to either a thermal phenomenon or the precession of a
large scale emissivity pattern (e.g. a spiral arm).  One exception is
Arp~102B, in which the relative fluxes of the red and blue peaks of
the profile varied sinusoidally with a period of 2 years (of the same
order as the dynamical time) over an interval of four years
\citep{ne97}. A very similar variability pattern, with the same
period, also appeared five years later \citep{gezari05}.  We also note
that in addition to gross profile variability that occurs on
timescales of years, some objects also exhibit sporadic variability on
timescales of only a few months; these variations may be related to a
phenomenon that occurs on the dynamical time scale.

The light crossing time of the disk of 3C~390.3 computed from
equation (\ref{tl}) is in the range of 11 to 37 days \citep[using the
inner and outer radii of the line-emitting disk from][]{eh94}.  This
affords an immediate test of the disk interpretation of the
double-peaked lines. A reverberation mapping campaign by the {\it
International AGN Watch} \citep{d98} yielded a lag between the
continuum and Balmer line variations of approximately 20~days, which
is in good agreement with the above value.

The Eddington ratios were computed by comparing the total bolometric
luminosity to the Eddington luminosity ($\ledd =
1.38\times10^{38}\;M/\msol\; {\rm erg\;s^{-1}}$). We used the
available Spectral Energy Distributions (SEDs) and integrated
bolometric luminosities for M~81 and NGC~4579 \citep{ho99} and
NGC~1097 \citep{n04,n05}. \citet{ehc03} presented an SED for Arp~102B,
and we adopt their SED, excluding the IR bump and adjusting the
specific luminosity in the X-ray band.  \footnotemark\footnotetext{
We adopted $\log[\nu\;L_{\nu}(1\;{\rm keV})]=42.331$ for consistency
with earlier observations with the {\it ROSAT} and {\it Einstein}
X-ray observatories \citep{h97,bier81}.  During the {\it ASCA}
observation \citep{ehc03} the X-ray flux was found to be a factor of 3
higher than the earlier observations and was attributed to an
unusually high state. We retained the 2--10~keV spectral slope
measured from the {\it ASCA} X-ray spectrum.} An SED for 3C~390.3 was
complied from various sources in the literature, and the full
references are provided in Table~\ref{results_tab}. There are no SEDs
for the remaining objects and we estimated their bolometric
luminosities using their measured X-ray luminosities and assuming a
value for $L_{2-10\;{\rm keV}}$/$L_{\rm bol}$, as given in
Table~\ref{results_tab}. For IRAS~0236.6--3101, 1E~0450.3--1817, and
NGC~4203; we adopted the average $L_{2-10 {\rm keV}}$/$L_{\rm bol}$
for the LINERs in this study (Arp~102B, NGC~1097, M~81, and
NGC~4579). For Pictor~A and PKS~0921--213 we adopted the radio-loud
quasar SED of \citet{e94} since this is the most likely SED for these
objects (had we adopted the LINER-like SED, even the lower limit to
the Eddington ratio would be larger than $10^{-2}$). To compute
$L_{bol}$, the SEDs were integrated from 1~MHz to 100~keV, using
power-law interpolations between data points.  We note that the
2--10~keV flux of an AGN is variable by a factor of a few, which
implies an additional uncertainty to the bolometric luminosities
estimated from the X-ray luminosity.

\begin{figure}
\epsscale{0.7}
\plotone{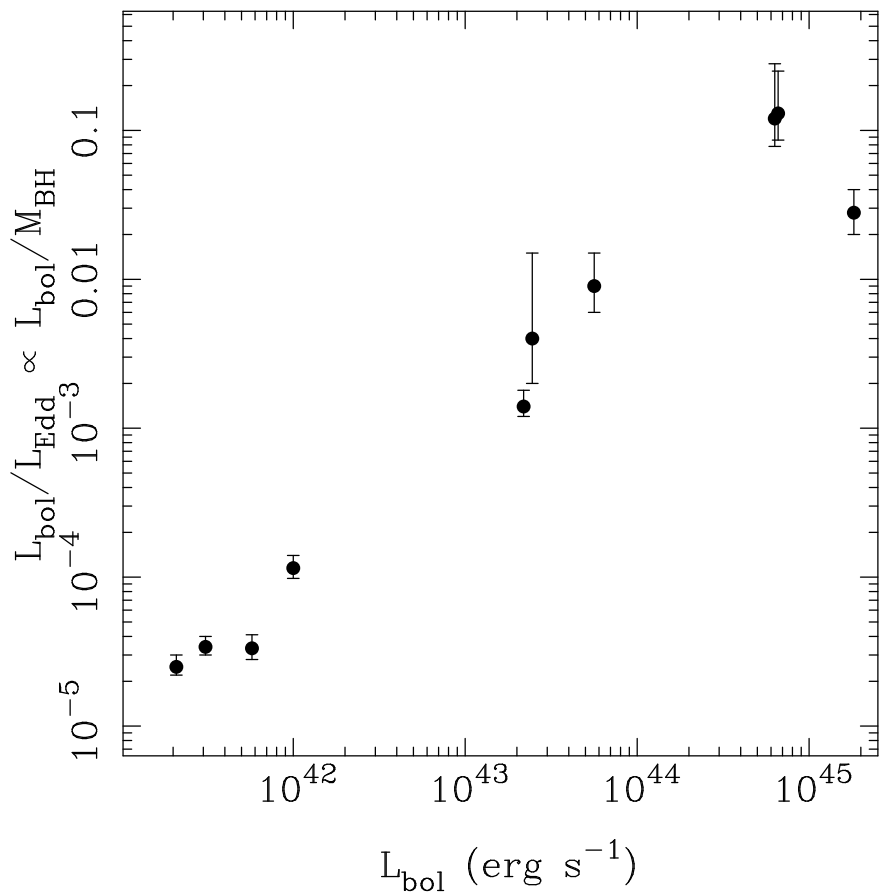}
\caption{\label{lbol_eddrat} Plot of Eddington ratio vs. the
bolometric luminosity. See Table \ref{results_tab} for the data and
\S\ref{mass_discussion} for details on how these quantities were
obtained. The apparent trend is very likely artificial since the
estimated black hole masses span only a factor of 10 (a factor of 2,
if 3C~390.3 is excluded), while the bolometirc luminosities span a
factor of $10^4$. Moreover, 3C~390.3 has $L_{bol}$ that is two orders
of magnitude larger than IRAS~0236.6--3101 and 1E~0450.4--1817, yet its
Eddington ratio is similar.}  
\end{figure}

The Eddington ratios listed in Table~\ref{results_tab} span a wide
range and there is a general positive trend between bolometric
luminosity and Eddington ratio, which we illustrate in
Figure~\ref{lbol_eddrat}. However, we do not consider this trend
convincing because the quantities plotted are not independent of each
other ($L_{bol}/\ledd\propto L_{bol}/M_{BH}$).  The trend is merely a
consequence of the fact that the estimated black hole masses span only
a factor of 10 (a factor of 2, if 3C~390.3 is excluded), while the
bolometirc luminosities span a factor of $10^4$. The existence of such
a trend is also called into question by the fact that 3C~390.3 has
$L_{bol}$ that is two orders of magnitude larger than
IRAS~0236.6--3101 and 1E~0450.4--1817, yet its Eddington ratio is
similar.

While many double-peaked emitters may harbor radiatively inefficient
accretion flows, the accretion flow in some objects, such as Pictor~A
and PKS~0921--213, are probably more similar to those found in Seyfert
1s. A vertically extended structure may still be required in these
objects to illuminate the outer accretion disk and may take the form
of a spherical corona such as that used by \citet{csd90}, a beamed
corona \citep{b99a,mbp01}, and/or the base of the jet \citep*{mff01}.

That the double-peaked emitters in even this small sample has a wide
range Eddington ratios is not completely surprising. The double-peaked
emitters found in the SDSS by \citet{s03} are very
heterogeneous. Although 12\% were classified as LINERs, many others
had properties which were indistinguishable from the general $z<0.332$
AGN population. We note that our results confirm the {\it general}
findings of \citet{wl04}, who estimated black hole masses for both the
\citet{eh94} and \citet{s03} samples through a series of correlations
obtained from reverberation mapping of AGNs \citep{k00}. These authors
found that (1) double-peaked emitters are a heterogeneous group with a
wide range in Eddington ratios and (2) those objects with higher
bolometric luminosities tend to have larger Eddington ratios.
However, the black hole masses of individual double-peaked emitters
deduced through that method can be very inaccurate, as acknowledged by
these authors. Although the black hole masses for 1E~0450.3--1817 and
Arp~102B inferred by these authors are consistent with those obtained
through the \msig\ relationship, the black hole masses for Pictor~A
and IRAS~0236.6--3101 were overestimated by an order of magnitude.
Thus we caution that for the purposes of modeling and interpreting
individual objects, it is necessary to obtain black hole masses
through a method which is more well tested and well calibrated in the
range of BH masses relevant to our objects, such as the
\msig~relationship.

\acknowledgments

K.T.L. was funded by the NASA Graduate Research Fellows Program
(NGT5-50387). We thank the CTIO staff for their expert help during the
observing run and the Association of Universities for Research in
Astronomy for financial support for travel to the observatory. We are
grateful to Aaron Barth, Steinn Sigurdsson, and the anonymous referee
for helpful discussions and suggestions. We thank A. Barth for
allowing us to use his spectra to test our fitting method.

\clearpage
\appendix
\section{Telluric Correction}\label{mass_ap1}

\begin{figure}
\epsscale{1.0}
\plotone{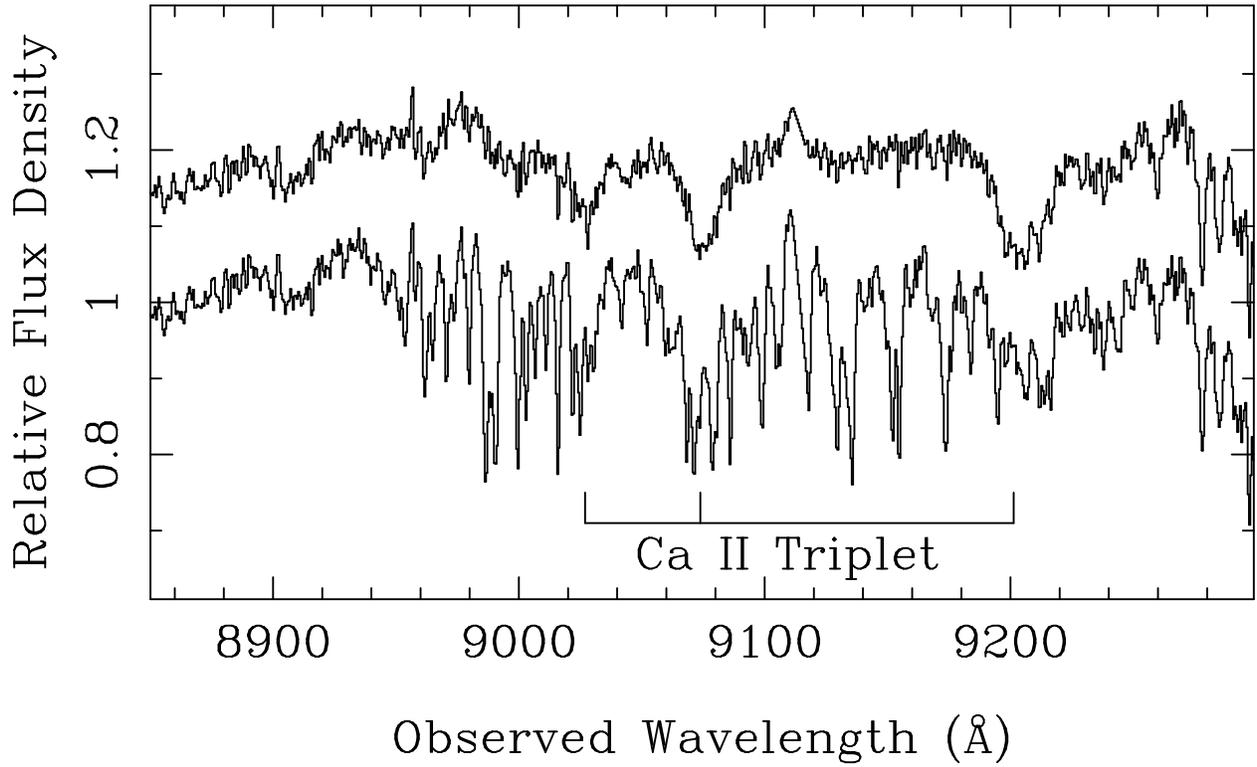}
\caption{\label{iras_fig} Average of the IRAS~0236.6--3101 spectra in
the absence of telluric correction. The positions of the \caii\ triplet
lines are marked. For comparison, the final spectrum obtained when the
telluric correction is performed is overplotted, with an arbitrary
vertical offset.}
\end{figure}

Removal of the telluric water vapor absorption lines was an essential
component of the data reductions because, with the exception of
NGC~1097, at least one of the \caii\ absorption lines in each target
galaxy was embedded in the telluric water vapor lines at wavelengths
longer than 8900~\ang.  To illustrate this we show in
Figure~\ref{iras_fig} the uncorrected spectrum of IRAS~0236.6--3101
in which all three \caii\ lines are embedded in the telluric water
vapor bands. To create a template of the atmospheric transmission, it
is common to observe an object with a nearly featureless continuum in
which any sharp features can be attributed to atmospheric
absorption. The atmospheric transmission changes not only with
position in the sky and airmass, but also throughout the
night. Because we wished to observe telluric standards several times
each night, we opted to observe rapidly rotating B-stars, with
rotational velocities in excess of 200~$\kms$.  Although these stars
are not as featureless as white dwarfs, they have the advantage of
being more common, allowing us to find a telluric standard in a
similar region of the sky as our targets.  More importantly, they are
very bright, requiring only $\sim 30$~s of exposure.

\begin{figure}
\epsscale{1.0}
\plotone{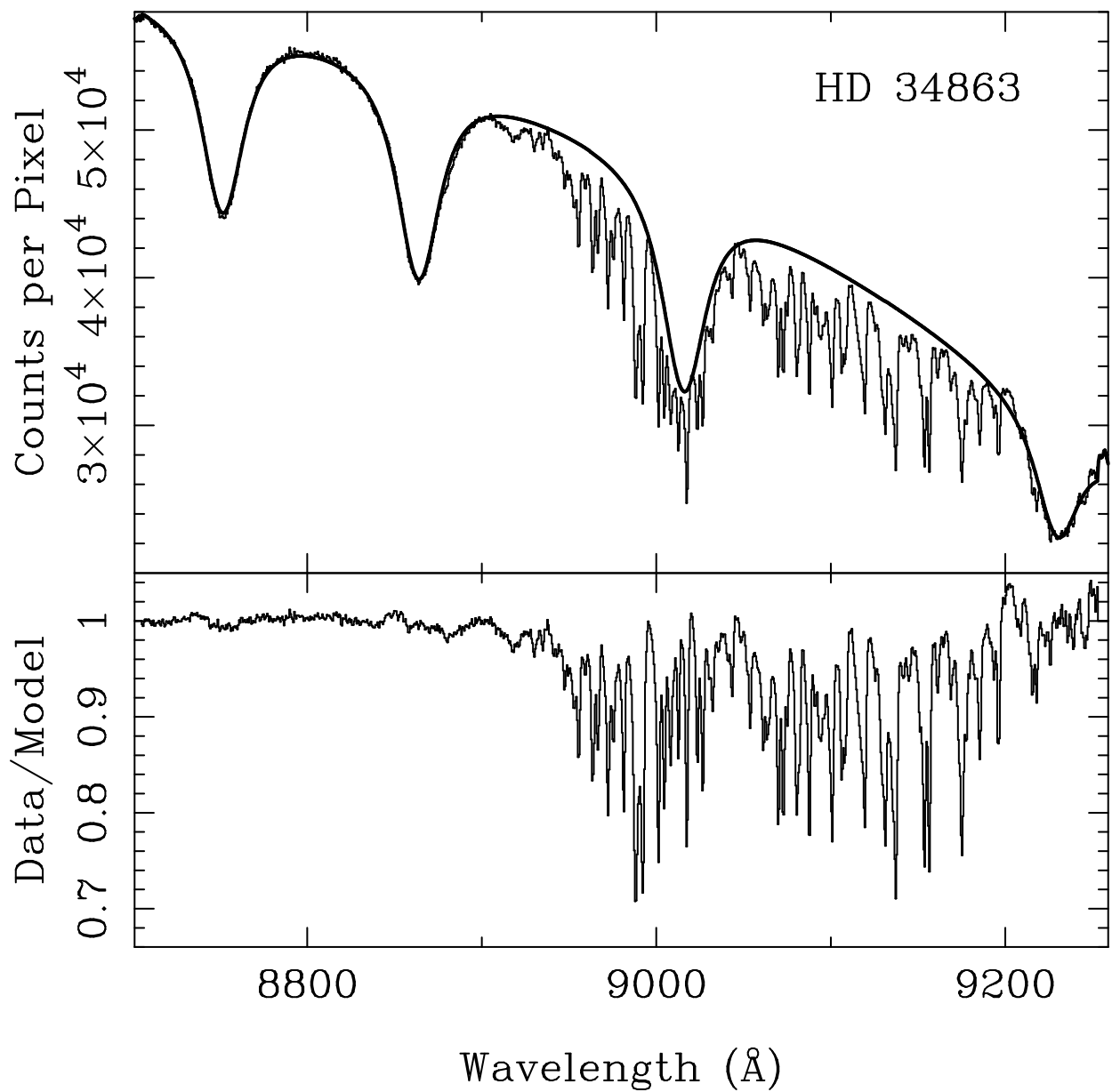}
\caption{\label{bstar} {\it Top:} Spectrum of the rapidly rotating
star HD~23683 (histogram) and the best-fitting model for its continuum
and hydrogen Paschen lines (solid line). {\it Bottom: } The residuals
to the fit, which are used as a template of the telluric water vapor
lines.  Before this template is applied to the galaxy spectra, the
values at $\lambda < 8900~\ang$ are set to unity and the rest of
the template is renormalized with a 1$^{\rm st}$ order polynomial.}
\end{figure}

Despite the rotational broadening, the Hydrogen Paschen lines were
still narrow enough that they could not be fitted as part of the
continuum as shown in Figure~\ref{bstar}. We modeled the intrinsic
spectrum of the B-star as a combination of a low-order ($n<4$)
polynomial and a set of Paschen absorption lines represented with
Voigt profiles. The best-fit Voigt profile parameters were determined
by fitting the interval from 8650 to 8875~\ang, a portion of the
B-star spectrum which had only three relatively un-blended Paschen
lines and was free from telluric absorption lines. Restricting the
profile parameters (except for their strengths) to narrow ranges
around the best-fit values, the B-star spectrum was then fitted over
the 8725--9300~\ang\ interval, which included four Paschen lines, two
of which were fitted in the previous step. A sequence of rejection
iterations was performed, in which the lower rejection level became
progressively more strict so as to force the fit to model the upper
envelope of the spectrum.  An example of this fit is shown in
Figure~\ref{bstar}. The B-star spectrum was then divided by this model
fit to generate a template of the telluric absorption lines and this
telluric template was set to unity at wavelengths less than 8900~\ang.

The residual template still exceeded unity near the red end of the
spectrum, particularly in the interval from 9195 to 9215$\ang$,
located on the blue wing of the 9229.75$\ang$ Pa$\zeta$ line. The
depth of this Paschen line was underestimated, because there was not
sufficient spectral coverage redward of the line to determine the true
continuum level (see Fig.~\ref{bstar}). A first-order polynomial was
fitted {\it post facto} to the envelope of the template to ensure that
all values were less than unity. Nevertheless, we suspect that the
telluric template is overestimated by as much as 4\% in the interval
from 9195-9215$\ang$ and we exercise caution when using the template
over this range of wavelengths. The only object affected by this
defect is IRAS~0236.6--3101, where the \caii~$\lambda$8662 line
appears to be too deep; the impact of this is discussed in
\S\ref{mass_notes}. If one plans to use telluric templates derived
from rapidly rotating B-stars near 9200$\ang$, it is adviseable to
have sufficient spectral coverage red-ward of the Pa$\zeta$ line so
that the continuum level (and thus the true line depth) can be
determined.

Of the six rapidly rotating B-stars observed, the simple Voigt
profile model described above only yielded a satisfactory fit
(i.e. residuals less than 1\%) in HD~34863. However, the humidity was
fortunately low and extremely stable over the course of the observing
run. When the template derived from HD 34863 was applied to the other
rapidly rotating B-stars, the residuals were less than 3\% in all
cases and sometimes less than 1\%, thus we were able to apply this
single template to all the galaxy spectra with little additional
degradation in the S/N.

The telluric absorption lines are not resolved in these data and it is
possible that the lines are so severely blended that they form a
pseudo-continuum that would not be corrected for by our
method. However, we have verified that this is not the case. A
spectrum of the telluric water vapor lines derived from a high
resolution ($\Delta\lambda/\lambda \sim 150,000$) near-IR spectrum of
Arcturus \citep{arcturus} was convolved by a Gaussian with a
dispersion of 1.35~$\ang$. There was no pseudo-continuum associated
with the convolved spectrum and there was excellent agreement between
the overall shape of the convolved telluric template and the telluric
template used in this paper.

The spectrum of HD 34863 was also used to derive a telluric template
to correct the spectra of the G and K giant stars from 8160 to 8400~\ang.
In this region, the B-star spectrum is featureless, and deriving the
template was straight forward. This template was applied to the stars
using the same method as for the galaxies.

\begin{deluxetable}{lccccc}
\tablecaption{\label{gal_obs}Galaxy Properties}
\tablewidth{5.5 in}
\tablehead{
  \colhead{Galaxy} & \colhead{}                     & \colhead{$f_{\nu}(8500~\ang$)}   & \colhead{Extraction}        & \colhead{Exposure} & \colhead{}\\
  \colhead{Name}   & \colhead{$z$\tablenotemark{\;\rm a}} & \colhead{(mJy)\tablenotemark{\;\rm b}} & \colhead{size ($''$, kpc)}  & \colhead{Time (s)} & \colhead{S/N\tablenotemark{\;\rm c}}
          }
\startdata
NGC~1097         & 0.0043 & 26.0 & 5.0 (0.45)   & 5400  & 115\\
Pictor A         & 0.035  &  3.0 & 4.5 (3.3)    & 18000 & 45\\
PKS 0921--213    & 0.0531 &  3.0 & 7.0 (7.7)  & 14400 & 40\\
1E~0450.3-1817   & 0.0616 &  0.5 & 5.0 (6.1)    & 18000 & 25\\
IRAS 0236.6-3101 & 0.0623 &  3.2 & 5.5 (7.1)    & 12600 & 80\\
\enddata
\tablenotetext{a\;}{Redshifts taken from \citet{eh04}}
\tablenotetext{b\;}{These fluxes were determined from spectra taken with a narrow slit and can thus be uncertain by up to a factor
of two.}
\tablenotetext{c\;}{S/N per pixel in the continuum regions in the vicinity of the \caii\ triplet
absorption lines.}
\end{deluxetable}

\begin{deluxetable}{lc|lc}
\tablecaption{\label{star_obs}Template Stars}
\tablewidth{3.5 in}
\tablehead{
  \colhead{Star} & \colhead{Spectral} & \colhead{Star} & \colhead{Spectral} \\
  \colhead{Name} & \colhead{Type}     & \colhead{Name} & \colhead{Type}
          }
\startdata
HD 3013   & K5 III & HD 21    & K1 III \\
HD 79413  & K1 III & HD 3809  & K0 III \\ 
HD 3909   & K4 III & HD 2224  & G8 III \\ 
HD 2066   & K3 III & HD 5722  & G7 III \\ 
HD 225283 & K2 III & HD 14834 & G6 III \\ 
\enddata
\end{deluxetable}

\begin{rotate}
\begin{deluxetable}{lcccccccccc}
\tablecaption{\label{results_tab}Velocity Dispersions and Derived Properties}
\tabletypesize{\footnotesize}
\tablewidth{8.5 in}
\tablehead{
  \colhead{Galaxy} & \colhead{$\sigma$}  & \colhead{$\mbh$}           & \colhead{$L_{\rm 2-10\,keV}$}   & \colhead{X-ray}                  &                                                  & \colhead{$L_X/L_{bol}$} &  \colhead{Range in}        & \colhead{$\tau_{dyn}$}  & \colhead{$\tau_{th}$}   & \colhead{$\tau_s$}\\
  \colhead{Name}   & \colhead{($\kms$)}  & \colhead{($10^{7}~\msol$)} & \colhead{(${\rm erg\;s^{-1}}$)} & \colhead{Ref.\tablenotemark{\;\rm d}}  & \colhead{$L_X/L_{bol}$}  &  \colhead{Ref.\tablenotemark{\;\rm e}}                & \colhead{$L_{bol}/\ledd$}  & \colhead{(months)}      & \colhead{(years)}       & \colhead{(years)}
          }
\startdata
NGC~1097          & 196$\pm$5                   & 12$\pm$2          & $4.4\times10^{40}$ & 1    & 0.09 & 1 & (2--3)$\times10^{-5}$   & 6--8   & 5--7   & 70--100 \\
Pictor A          & 145$\pm$20                  & 4$\pm$2           & $4.0\times10^{43}$ & 2,3  & 0.07 & 2 & 0.08--0.3               & 1--4   & 1--3   & 15--40  \\
PKS 0921--213     & 144$^{\;+18}_{\;-16}$       & 4$\pm$2           & $4.2\times10^{43}$ & 4    & 0.07 & 2 & 0.09--0.3               & 1--4   & 1--3   & 15--40  \\ 
1E~0450.3--1817   & 150$^{\;+30}_{\;-25}$       & 4$^{\;+4}_{\;-3}$ & $3.2\times10^{42}$ & 5    & 0.13 & 3 & (0.2--2)$\times10^{-2}$ & 0.5--5 & 0.5--4 & 10--55   \\
IRAS 0236.6--3101 & 154$\pm$15                  & 5$\pm$2           & $7.3\times10^{42}$ & 6    & 0.13 & 3 & (0.6--2)$\times10^{-2}$ & 2--4   & 2--3   & 35--50  \\
\cline{1-11}
Arp~102B          & 188$\pm$8 \tablenotemark{\;\rm a} & 11$\pm$2          & $6.3\times10^{42}$ & 7    & 0.17 & 4 & (1--2)$\times10^{-3}$   & 5--8   & 4--7     & 60--90  \\
3C 390.3         & 273$\pm$16\tablenotemark{\;\rm b} & 50$\pm$10         & $1.9\times10^{44}$ & 3    & 0.11 & 5 & (2--4)$\times10^{-2}$   & 20--40 & 17--30   & 230-430  \\
NGC~4579          & 165$\pm$4 \tablenotemark{\;\rm a} & 6$\pm$1           & $2.0\times10^{41}$ & 8    & 0.20 & 6 & (1--2)$\times10^{-4}$   & 3--4.5 & 2.5--4   & 35--50  \\
NGC~4203          & 167$\pm$3 \tablenotemark{\;\rm a} & 6$\pm$1           & $4.0\times10^{40}$ & 9    & 0.13 & 3 & (3--4)$\times10^{-5}$   & 3--4.5 & 2.5--4   & 35--50  \\
M81               & \dotfill             & 6$\pm$1\tablenotemark{\;\rm c} & $1.9\times10^{40}$ & 8    & 0.09 & 6 & (2--3)$\times10^{-5}$   & 3--4   & 2.5--4   & 35--50  \\
\enddata
\tablenotetext{a\;}{Velocity dispersions taken from \citet{b02}.}
\tablenotetext{b\;}{Velocity dispersion taken from \citet{nel04}.}
\tablenotetext{c\;}{Black hole mass taken from \citet{b00} and \citet*{dev03}; derived from spatially-resolved stellar
and gas kinematics, respectively.}
\tablenotetext{d\;}{{\it References for X-ray luminosity and spectra.}
  -- (1) \citet{n04} and \citet{n05}; (2) \citet{eh98}; (3) \citet*{sem99}; (4)
  \citet{l06}; (5) \citet{stocke83}; (6) \citet{boller92}; (7)
  \citet{ehc03}; (8) \citet{ho99}; (9) \citet{i98}.}
\tablenotetext{e\;}{{\it References for SED and $L_{\rm 2-10\;keV}/L_{bol}$.}
-- (1) \citet{n04} and \citet{n05}; (2) assumed radio-loud QSO SED of \citet{e94};
(3) assumed average $L_{X}/L_{bol}$ for the LINERs in this study; (4) SED from 
compilation in \citet{ehc03} (without the IR bump) except the X-ray flux 
density was scaled to match the ROSAT 0.5--2.0 keV flux \citet{h97} (5) 
SED compiled from \citet[][radio]{alef88}, \citet[][IR]{m84},
\citet[][optical]{vz91}, \citet[][; UV]{z96}, \citet[][X-ray]{mbc94};
(6) \citet{ho99}. For more details, see \S\ref{mass_discussion}.}
\end{deluxetable}
\end{rotate}

\begin{rotate}
\begin{figure}
\epsscale{1.2}
\plotone{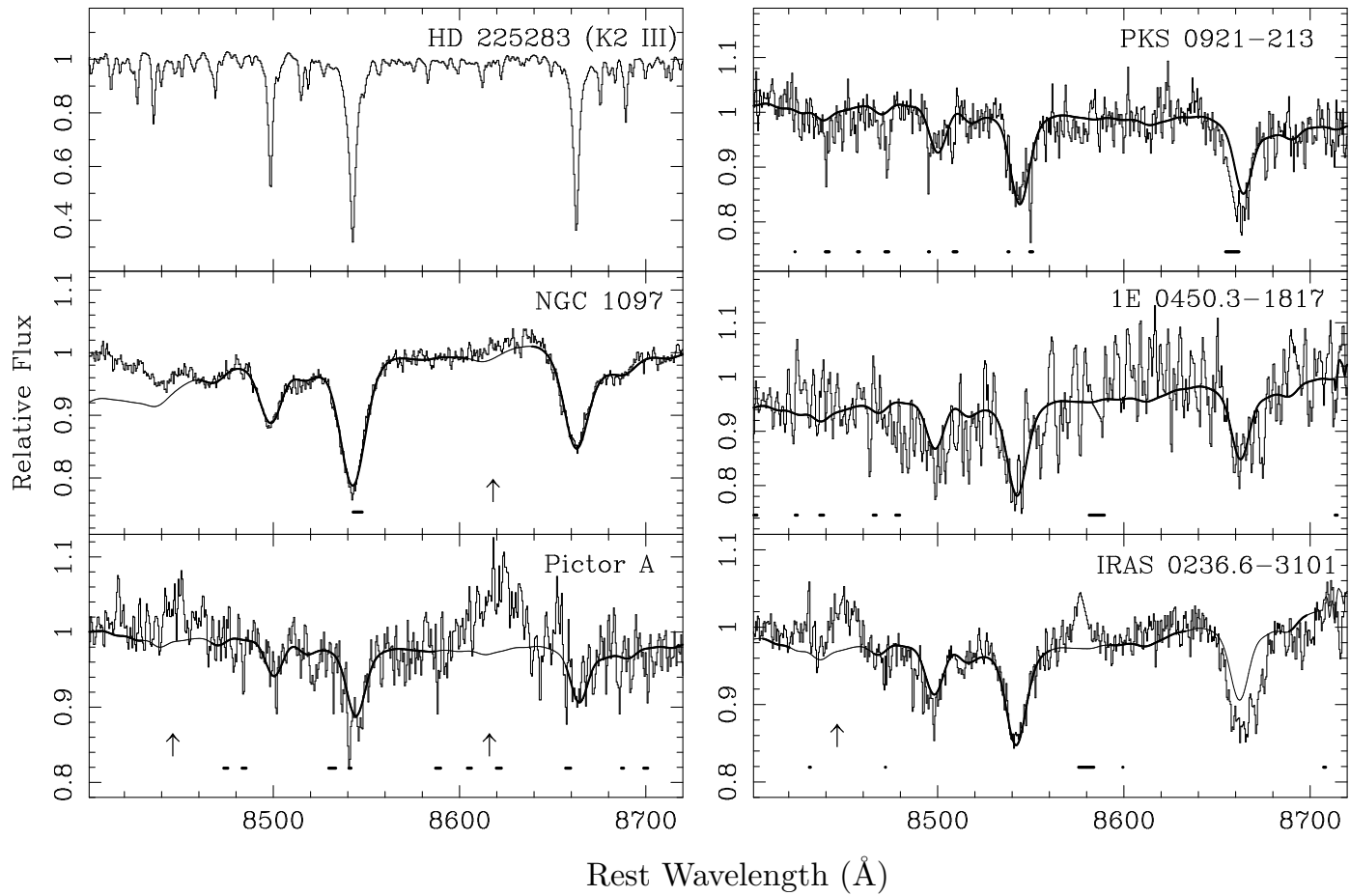}
\caption{\label{spectra} The \caii\ spectral region in a template star
and the five target galaxies. The best-fitting model for each galaxy
is overplotted and in the intervals used to perform the fit, the model
is overplotted with a thick solid line.  Data points with unusually
large error bars --- due to a strong night sky line, a cosmic ray, or
a bad column --- are indicated with black dots along the bottom of the
spectrum.  The \ion{O}{1}~$\lambda8446$ and
[\ion{Fe}{2}]~$\lambda8618$ emission lines, when present, are
indicated by an arrow.}
\end{figure}
\end{rotate}

\end{document}